\documentstyle[amssymb,preprint,aps,epsfig]{revtex}
\tightenlines

\begin{document}
\title{Revivals and oscillations of the momentum of light in a planar multimode
waveguide}
\author{Yuri B. Ovchinnikov and Tilman Pfau}
\address{5. Physikalisches Institut, Universit\"{a}t Stuttgart, Pfaffenwaldring 57,\\
D-70550 Stuttgart, Germany}
\date{\today}
\maketitle

\begin{abstract}
The evolution of the transverse momentum of monochromatic light entering a
multimode planar waveguide at large angle is investigated. We report on
oscillations of the momentum caused by the beatings between the adjacent
populated modes of the waveguide and their periodic collapses and revivals.
A new type of an interferometer based on this effect with fringe spacing as
small as $\lambda /9$ is demonstrated experimentally and periods as small as
$\lambda /1000$ seem to be feasible.
\end{abstract}



The propagation of light along a planar multimode waveguide is one of the
fundamental questions of wave optics. This problem is closely related to the
evolution of atomic wavepackets in square potential wells \cite{Kaplan}. As
it was shown in \cite{Bryngdahl}, \cite{Ovchinnikov} it is also related to
the propagation of light in a periodic grating. The situation, when the
monochromatic light enters the waveguide with no average transverse
momentum, was studied in \cite{Ovchinnikov}, \cite{Ulrich}.

This letter considers the case, when the monochromatic light beam enters the
planar metal multimode waveguide at some large angle. In this case the light
experiences not only full and fractional revivals \cite{Ovchinnikov} of its
initial phase distribution along the waveguide, but also additional
oscillations of the transverse momentum. This is related to the beatings
between the adjacent populated transverse modes of the waveguide. These
oscillations are similar to the Pendell\"{o}sung oscillations \cite{Ewald}
for the Bragg reflection of waves at diffraction gratings, which were
observed in many different systems \cite{Shull}, \cite{Kunze}.

Based on this effect, a new type of interferometer with a fringe spacing
much smaller than the wavelength of light has been demonstrated.

The basic setup is shown in Fig.\thinspace 1. The waveguide is realized by
two metal mirrors placed parallel to each other at distance $d$. The plane
monochromatic light wave enters the waveguide at an angle $\beta $. The
amplitude of the transverse distribution of the light field at the front of
the waveguide is $E_{0}(x,y<0)=E_{0}\exp \left( -ik\sin \left( \beta \right)
x\right) \exp \left( -ik\cos \left( \beta \right) y\right) $, where $%
k=2\pi/\lambda $ and $\lambda $ is the wavelength of the light. Therefore,
the initial distribution\ has a constant amplitude and its phase changes
linearly in $x$ direction. The modes propagating along the waveguide have to
satisfy the condition of quantization of the transverse component of their
wavevector
\begin{equation}
kd\sin \alpha {_{\text{n}}}=\text{n}\pi ,  \label{1}
\end{equation}
where n=1,2,3... and $\alpha _{\text{n}}$ is the angle of the wavevector in
the n-th mode with respect to the boundaries of the guide. The field
amplitude inside the waveguide is given by
\begin{equation}
E(x,y)=\sum_{n=1}^{\infty }c_{\text{n}}\sin {\ \left( \frac{n\pi }{d}%
x\right) }\exp {\ \left( -ik\sqrt{1-\left( \frac{\text{n}\pi }{kd}\right)
^{2}}y\right) }\exp {\ \left( -\gamma _{\text{n}}y\right) }.  \label{2}
\end{equation}
The exponents give the phase and the loss of the n-th mode along the
waveguide and $\sin {\left( \text{n}\pi x/d\right) }$ - is the distribution
of the light amplitude in the n-th mode. The coefficients $c_{\text{n}}$ are
determined by the projection of the initial distribution $E_{0}(x,y<0)$ onto
the modes of the waveguide
\begin{equation}
c_{\text{n}}=\frac{2}{d}\int_{0}^{d}\sin {\ \left( \frac{n\pi }{d}x\right) }%
E_{0}\left( x,y<0\right) dx,  \label{3}
\end{equation}
which can be integrated analytically. The transverse momentum distribution
of the light emerging from the waveguide can be found by the Fourier
transform
\begin{equation}
E(k_{x})\propto \frac{1}{d}\int_{0}^{d}E(x,y=L)\exp {\left( -ik_{x}x\right) }%
dx,  \label{4}
\end{equation}
of the spatial distribution of the light amplitude $E(x,L)$ at the output of
the waveguide, where $L$ is the length of the waveguide and $k_{x}$\ is the
transverse component of the light wavevector.

The numerical calculations of the light momentum distribution at the output
of the waveguide based on equations (2-4) show quasi-periodic oscillations
between the two main directions $\beta $ and $-\beta $ (see Fig.\thinspace
1) as a function of the length $L$ of the waveguide. We shall label these
two output beams respectively as transmitted and reflected beams. For
incidence angles $\beta >\arcsin \left( \lambda /d\right) $ the two output
beams are well separated from each other in momentum.\ The solid curve in
Fig.\thinspace 2 shows the calculated intensity of the transmitted light
beam as a function of the waveguide length $L$. In these calculations we
assumed large width of the waveguide $d>>\lambda $ and small angles of
incidence $\sin \left( \beta \right) <<1$. We used a dimensionless relative
length $s=L/L_{r}$, corresponding to the number of revival, or Talbot, or
self imaging periods \cite{Ovchinnikov}

\begin{equation}
L_{r}=\frac{8d^{2}}{\lambda }
\end{equation}
and a constant propagation parameter $f=48$,
\begin{equation}
f=\frac{L_{r}}{L_{b}}=\frac{4d\sin \left( \beta \right) }{\lambda },
\end{equation}
which is equal to the ratio of the revival period $L_{r}$ to the period of
light momentum beatings

\begin{equation}
L_{b}\simeq \frac{2d}{\sin \left( \beta \right) }.
\end{equation}
This is related to the situation when the waveguide width $d$ is fixed and
its length $L$ is varied. For this setting there are only a few transverse
modes of the waveguide significantly populated, which satisfy the condition $%
{\alpha }_{n}^{e}\simeq {\beta }$. The length $L_{b}$ corresponds to the
period of beatings between these modes. The expression (7) is valid only for
$d>>\lambda $ and $\sin \left( \beta \right) <<1$, which is the case for our
experiment. Without this assumption the period of the beatings between two
adjacent modes is
\begin{equation}
L_{b}=\lambda /\left( \sqrt{1-\sin ^{2}\alpha _{n}^{e}}-\sqrt{1-\sin
^{2}\alpha _{n+1}^{e}}\right) .
\end{equation}
Note that the period (7) can be found also from a zigzag geometrical ray
propagation of the ingoing light beam through the waveguide. The
corresponding classical oscillations of the intensity in the transmitted
beam are shown in Fig.\thinspace 2 by the dashed sinusoidal trace possessing
a period $L_{b}$. However these classical oscillations do not depend on the
light wavelength.

The real wave behavior (solid curve) is quite different and can be
characterized by two main periods $L_{r}$ and $L_{b}$. Despite the fact,
that the period of the momentum beatings is approximately the same as in the
classical case, the location of the peaks is determined by the position of
revival resonances \cite{Ovchinnikov} and therefore depends on the
wavelength of light. The amplitude of the beatings is modulated in such a
way that they are maximal around the relative lengths of the waveguide $%
s_{m}^{\max }=m/2$ and minimal around $s_{m}^{\min }=1/4+m/2$, with $%
m=0,1,2...$. It means the revivals and collapses of momentum state\
oscillations take place. For full revival lengths, when $m$ is even and $%
s_{m}^{\max }=0,1,2...$, the revivals of the initial light momentum take
place and all light is concentrated in the transmitted output beam. For
half-revival lengths, when $m$ is even and $s_{m}^{\max }=1/2,3/2,5/2...$,
the phase of the initial momentum distribution is inverted and all light is
concentrated in the reflected beam. It can be seen from Fig.\thinspace 2
that the phase of momentum oscillations at $s=1/2$ is inverted with respect
to their phase at $s=0$, which is not the case for classical oscillations.
From equations (5-7) it follows that the tuning of the waveguide from one
transmission fringe to the next can be done by changing either the length or
the width of the waveguide, or the light wavelength according to

\begin{equation}
\frac{\delta L}{L}=\frac{2\delta d}{d}=\frac{\delta \lambda }{\lambda }=%
\frac{L_{b}}{L}\simeq \frac{1}{sf}=\frac{2d}{L\sin \left( \beta \right) }.
\label{8}
\end{equation}

One way to change the relative length $s$ of the waveguide is to vary the
width $d$, while the length of the waveguide $L$ fixed. As $s=L\lambda
/8d^{2}$ is nonlinear in $d$, the fringe spacing $\delta d$ also changes
nonlinearly in $d$ as
\begin{equation}
\delta d=\frac{d^{2}}{L\sin \left( \beta \right) }.
\end{equation}
The maximum amplitudes of the momentum oscillations are located at half and
full revival distances, which are characterized by the following widths of
the waveguide

\begin{equation}
d_{m}=\sqrt{\frac{L\lambda }{8s_{m}^{\max }}}=\sqrt{\frac{L\lambda }{4m}},
\label{5}
\end{equation}
where $m$ is an integer.

Compared to all other types of interferometers, where the minimal fringe
spacing is equal to $\lambda /2$, the distance $\delta d$\ can be much
smaller than the light wavelength. This can be explained by the fact, that
in a waveguide the interfering modes experience multiple reflections between
the mirrors before their phase difference is observed.

Experimentally the waveguide was formed by two parallel flat mirrors of
equal length ($L=5$ cm), coated with bare gold. The substrates were made of
fused silica and polished to a surface figure $\backsim \lambda /20$. All
degrees of freedom of one of the mirrors were adjustable. This mirror was
attached to a precise mechanical translational stage driven by a
''COHERENT'' Encoder Driver System 37-0486 which provided a resolution in
translation of $\backsim 10$ nm and a total range of 13$\,$mm in the x
direction.

We used a He-Ne laser with wavelength $\lambda =633$ nm, divergence about $%
1\,$mrad, linear polarisation, directed parallel to the mirrors, and a beam
diameter of $\simeq 2\,$mm. In our first experiment the entrance edge of the
waveguide was illuminated at an angle of $\beta =0.253$ rad.

To detect the oscillations we used a photodiode, which was placed in the
reflected output light beam (Fig.\thinspace 1) at a distance $D\simeq 10\,$%
cm behind the waveguide. Each measurement of the light intensity in the
reflected beam as a function of the waveguide width $d$ was done in a single
sweep of the driven mirror. The corresponding dependencies are shown in
Fig.\thinspace 3. We have observed quasi-periodic oscillations of the light
intensity in the reflected and the transmitted output light beam, shifted by
$\pi $ with respect to each other. This leads to the conclusion, that the
output light momentum experiences quasi-periodic oscillations between the
two directions $\beta $ and $-\beta $. Fig.\thinspace 3 also shows that
these oscillations experience quasi-periodic collapses and revivals as a
function of distance $d$. In full agreement with equation (10) the maximum
amplitudes of the oscillations were observed at $d_{1}\simeq 89\mu $m, $%
d_{2}\simeq 63\mu $m, $d_{3}\simeq 51.5\mu $m, $d_{4}\simeq 44.5\mu $m, $%
d_{5}\simeq 40\mu $m, $d_{6}\simeq 36\mu $m, $d_{7}\simeq 33.5\mu $m, $%
d_{8}\simeq 31.5\mu $m and $d_{9}\simeq 29.7\mu $m, where half or full
revivals take place.

The amplitude of the fringes become smaller with decreasing the width $d$ of
the waveguide, which has several reasons. First, by extinction of the total
light power coupled to the waveguide; second, by increasing the number of
reflections inside the waveguide and the corresponding losses; third, for
smaller width of the waveguide the divergence of the outgoing light beams
becomes larger and the intensity decreases. The period of fringes is
decreasing with reducing the width $d$ of the waveguide in full accordance
with equation (9). For $m=9$ ($d\simeq 30\mu $m) we have observed fringes
with a period $\delta d\simeq 70$ nm. These fringes are shown in the inset
of Fig.\thinspace 3c. For such a small fringe spacing the single steps of
the Encoder Driver, which are of the order of $10$ nm, were resolved.

To observe directly the oscillations of the output light between the two
directions $\beta $ and $-\beta $\ we used a CCD camera placed at distance $%
D\simeq 12\,$cm behind the waveguide, which detected the transmitted and
reflected output beams simultaneously. The detected transverse spatial
distribution of light around the $m=1$ revival resonance are presented in
the left column of Fig.\thinspace 4. The right column of the Fig.\thinspace
4 shows corresponding momentum distributions calculated numerically from
equations (2), (3) and (4).

\ According to the equation (10) the position of fringes depends on the
light wavelength. To demonstrate this fact experimentally, we have
illuminated the waveguide with two co-propagating and well overlapped laser
beams of different frequency ($\lambda _{1}=633$ nm and $\lambda _{2}=532$
nm). For the latter wavelength we used a frequency-doubled Nd:Vanadate
single-frequency laser. At large widths $d$ of the waveguide the outgoing
light distributions are identical for both frequencies, but for small $d$,
the fringes are shifted with respect to each other. As an example
Fig.\thinspace 4c shows spatial distributions of the output light at width $%
d=86\mu $m. Under these conditions the red light intensity ($\lambda
_{1}=633 $ nm) is almost completely contained in the transmitted beam and
the green light ($\lambda _{2}=532$ nm)\ in the reflected one. Therefore the
two different frequency components are well separated in momentum and space.

From all these observations we come to the conclusion, that the multimode
waveguide at large incidence angles of the input light beam can be
considered as a new kind of an interferometer. The definite advantage of
such an interferometer, compared to other known types, like Fabry-Perot or
Michelson interferometers, is that its fringe spacing (5) can be much
smaller than $\lambda /2$. The observed period of $\lambda /9$ was soley
limited by technical factors in our first experiment. For a very narrow
waveguide ($d\sim \lambda $) the period of revivals and mode beatings become
smaller than in the multimode waveguide considered above ($d>>\lambda $).
Especially interesting is the limit, when the waveguide contains only two
propagating modes. The fringes of such a waveguide becomes uniform and
periodic, and their period can be calculated from the formula (8), where $n=1
$ should be used. The fringe spacing of such an interferometer is expected
to be very small. For example in a case of a waveguide with $d=$ $0.7\,\mu $%
m, $L=100$ $\mu $m, $\lambda =633$ nm and $\beta =33^{\circ }$, the fringe
spacing is expected to be $\delta d=0.56$ nm.

Such a new kind of interferometer, especially in dielectric multimode
waveguides, can find many applications in precision optical measurements and
also in switching and modulation of optical signals.

We are very grateful to David Wilmering from the NIST in Gaithersburg for
producing the gold mirrors for this experiment and to Robert L\"{o}w for his
comments on this manuscript.

\newpage

\begin{figure}[!hbp]
\begin{center}
\epsfig{file=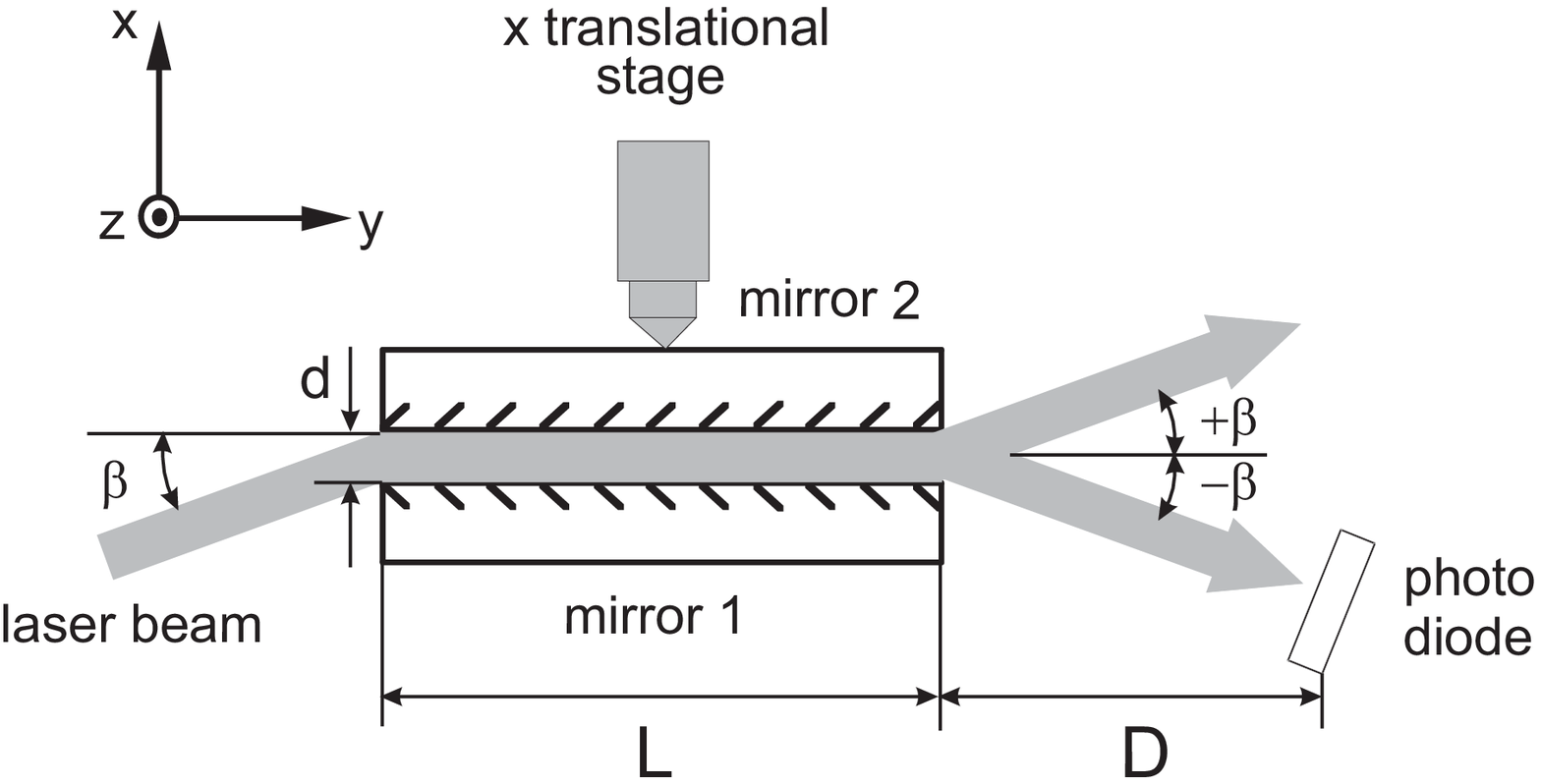, width=0.8\textwidth} \caption{Experimental
setup of the planar metal waveguide for observing the oscillations
and revivals of the transverse momentum of light.}
\end{center}
\end{figure}

\begin{figure}[!b]
\begin{center}
\epsfig{file=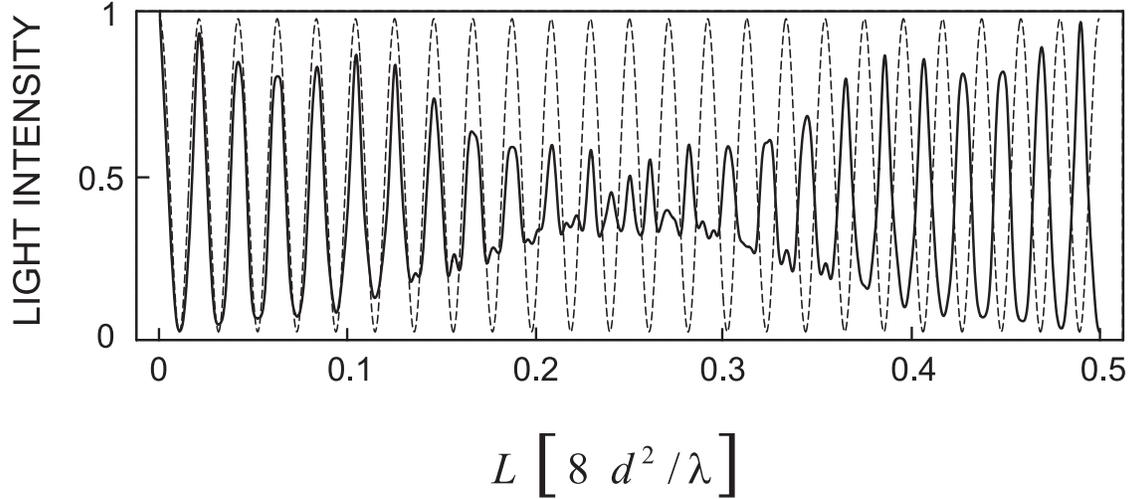, width=0.9\textwidth} \caption{Light
intensity of the transmitted output beam of the waveguide as a
function of its relative length, measured in revival periods
$L_{r}=8d^{2}/\lambda$. In this calculation the propagation
parameter $f=48$ was used. The influence of several waveguide
modes leads to a wavelength dependent deviation from the geometric
optics case for $L>L_r/8$.}
\end{center}
\end{figure}

\begin{figure}[!hbp]
\begin{center}
\epsfig{file=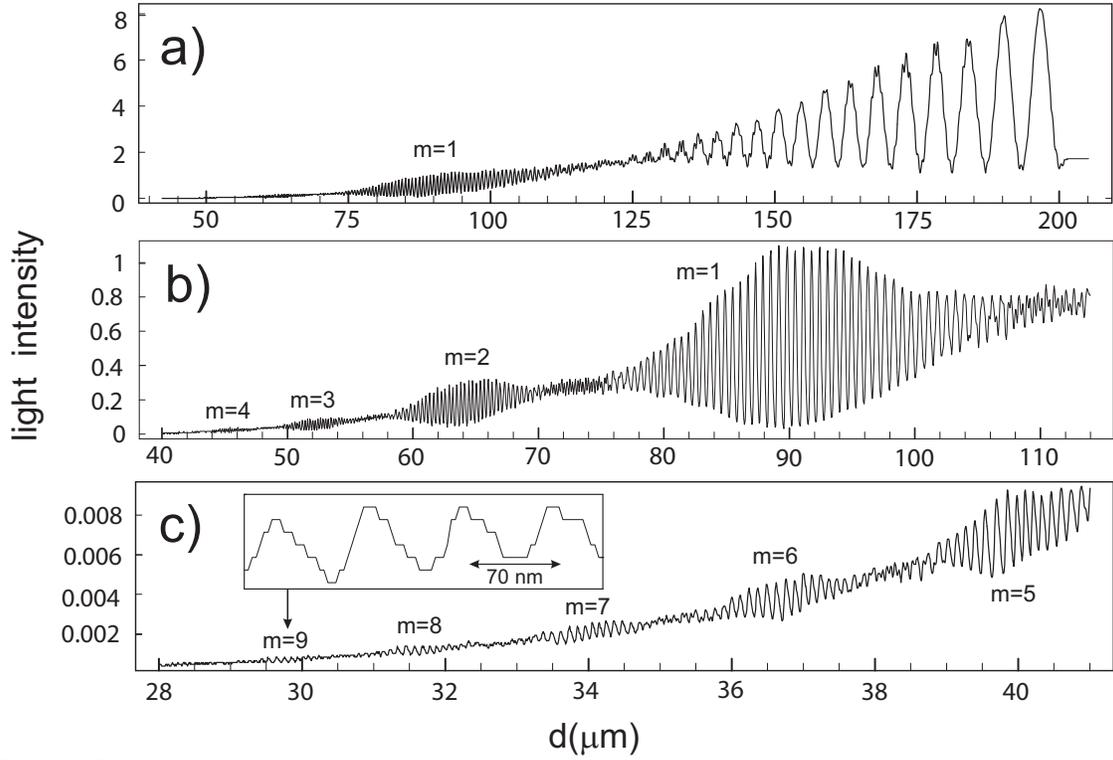, width=0.9\textwidth} \caption{Light
intensity in the reflected beam of the light past the waveguide as
a function of the waveguide width $d$.}
\end{center}
\end{figure}

\begin{figure}[!hbp]
\begin{center}
\epsfig{file=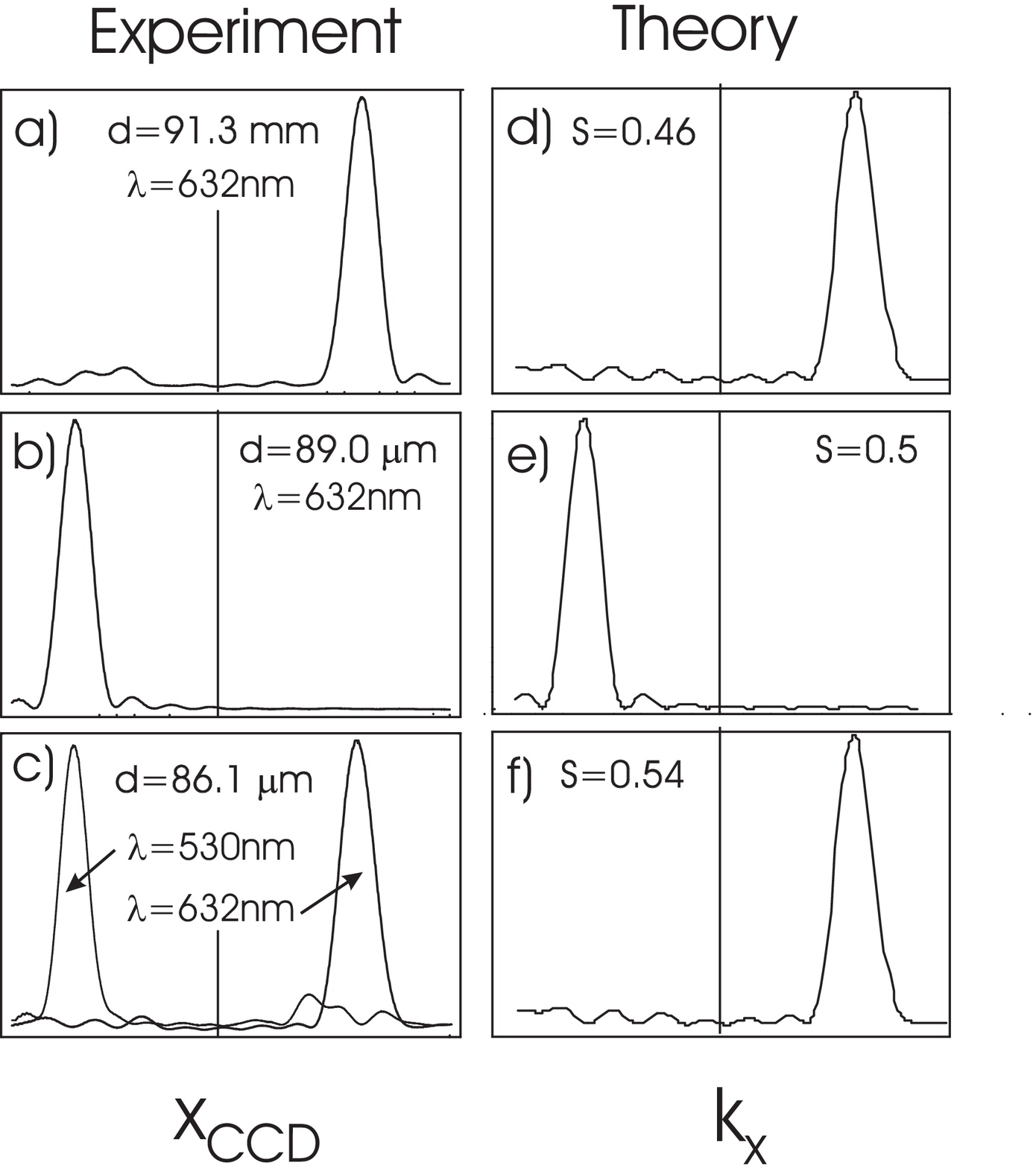, width=0.8\textwidth} \caption{a), b), c) -
experimental profiles of the light distributions in a far zone
behind the waveguide; d), e), f) - numerically calculated
transverse momentum distribution of light behind the waveguide.}
\end{center}
\end{figure}

\end{document}